\documentclass[3p,onecolumn,times]{elsarticle}
\usepackage[usenames]{color}
\usepackage{graphicx}
\usepackage{amssymb}
\usepackage{amsmath}
\usepackage{epstopdf}

\journal{Communications in Nonlinear Science and Numerical
Simulation}

\begin{document}
\biboptions{sort&compress}

\begin{frontmatter}

\title{New classes of solutions in the Coupled $\cal{PT}$ Symmetric Nonlocal
Nonlinear Schr\"odinger Equations with Four Wave Mixing}

\author[uae]{P.~S.~Vinayagam}
\address[uae]{Department of Physics,
United Arab Emirates University, P.O.Box 15551, Al-Ain, United
Arab Emirates}
\author[kmu]{R.~Radha}
\address[kmu]{Centre for Nonlinear Science (CeNSc), PG and Research Department of Physics, Government College for Women (Autonomous), Kumbakonam 612001, India}
\author[uae]{U.~Al~Khawaja\corref{cor1}}
\ead{u.alkhawaja@uaeu.ac.ae}
\cortext[cor1]{Corresponding author}
\author[china]{Liming Ling}
\address[china]{Department of Mathematics, South China University of Technology, Guangzhou, 510640, China}

\begin{abstract}
We investigate generalized nonlocal coupled nonlinear
Schor\"odinger equation containing Self-Phase Modulation,
Cross-Phase Modulation and Four Wave Mixing involving nonlocal
interaction. By means of Darboux transformation we obtained a
family of exact breathers and solitons including the Peregrine
soliton, Kuznetsov-Ma breather, Akhmediev breather along with all
kinds of soliton-soliton and breather-soltion interactions. We
analyze and emphasize the impact of the four-wave mixing on the
nature and interaction of the solutions. We found that the
presence of Four Wave Mixing converts a two-soliton solution into
an Akhmediev breather. In particular, the inclusion of Four Wave
Mixing results in the generation of a new solutions which is
spatially and temporally periodic called ``Soliton (Breather)
lattice".
\end{abstract}


\end{frontmatter}

\section{Introduction}
Since the invention of the laser, optical solitons \cite{hasegawa}
have played an important role in nonlinear physics. The optical
soliton in fibres is probably the best studied form of solitons
because of its remarkable behavior that agrees well with
theoretical predictions and its potential as optical information
carrier. The propagation of optical pulses through optical
birefringent fibres is described by the celebrated Manakov model
of the following form \cite{manakov1},
\begin{subequations}
\begin{eqnarray}
\textit{i} q_{1t} &+& q_{1xx} + 2 \left(g_{11}|q_1|^2 + g_{12}|q_2|^2\right)
q_1 = 0,\\
\textit{i} q_{2t} &+& q_{2xx} + 2 \left(g_{21}|q_1|^2 + g_{22}|q_2|^2\right)
q_2 = 0,
\end{eqnarray}\label{Manok}
\end{subequations}
where, $q_1$ and $q_2$ are wave envelopes, $x$, $t$ are space and
time variables and $i$ is the imaginary unit. The interaction
coefficients $g_{11}$ and $g_{22}$ correspond to the Self-Phase
Modulation (SPM) and $g_{12}$ and $g_{21}$ represent the
Cross-Phase Modulation (XPM) \cite{agrawal}. Eq.~(\ref{Manok}) has
been shown to be integrable if either (i) $g_{11}$ = $g_{12}$
=$g_{21}$ =$g_{22}$ or (ii) $g_{11}$ = $g_{21}$ =-$g_{12}$
=-$g_{22}$ \cite{pre-psv}. The first choice corresponds to Manakov
model \cite{manakov2} while the second choice represents the
modified Manakov model \cite{modifiedmanakov}. The above coupled
nonlinear Schr\"odinger equation (CNLSE) involving local nonlinear
interactions has been extensively studied in diverse fields such
as nonlinear optics \cite{nlo}, bio-physics \cite{bio}, finance
\cite{finance} and oceanographic studies \cite{oceano} etc.

After the discovery of new integrable model named ``nonlocal integrable NLSE "
by Ablowitz and Mussilimani \cite{ablow}, the integrable models involving
\textit{nonlocal interactions} have attracted considerable attention. In
ref.\cite{ablow} Ablowitz {\it et al}., introduced the model of the following
form
\begin{align}\label{ablowitz model}
\textit{i} q_t(x,t)= q_{xx}(x,t)+2\,g\,q(x,t)q^{\ast}(-x,t)\,q(x,t).
\end{align}
This equation (\ref{ablowitz model}) is $\cal{PT}$ symmetric in
the sense that the equation brings a self-induced potential of the
form $V(x,t)=q(x,t)q^{\ast}(-x,t)$ and satisfies the $\cal{PT}$
symmetric condition $V(x,t)=V^{\ast}(-x,t)$. It is \textit{
nonlocal} in the sense that the evolution of the field at
transverse coordinate $x$ always requires the information from the
opposite point $-x$ \cite{ablow}. Interestingly, in this study
Ablowitz \textit{et al.,} have shown the new model given by
Eq.~\eqref{ablowitz model} to be fully \textit{integrable} since
it possesses linear Lax pair and infinite number of conserved
quantities. Unlike the integrable model considering local
nonlinear interaction, the above new integrable  model involving
nonlocal interactions is very rich in the sense of giving rise to
new results and possessing some new behaviors, e.g., it
simultaneously admits both the bright and dark soliton solutions
for the same nonlinearity \cite{aksarma}. In addition, several
studies involving nonlocal $\cal{PT}$ symmetric optics can lead to
alternative classes of optical structures and devices with unique
properties. These include the effect of nonlinearity on beam
dynamics in $\cal{PT}$ symmetric potential \cite{exp1}, solitons
in dual-core waveguides \cite{exp2,exp3}, and Bragg solitons in
$\cal{PT}$ symmetric potentials \cite{exp4}. Finally $\cal{PT}$
symmetric concepts have also been studied in plasmonics
\cite{plasmonics}, optical metamaterials
\cite{opt-meta-1,opt-meta-2} and coherent atomic medium
\cite{cohe-atom}.

Four wave mixing (FWM) is one of the most important nonlinear
phenomena having practical applications, particularly in nonlinear
optics \cite{agrawal,yang} such as optical processing
\cite{opt-proc}, real time holography \cite{holo}, Phase conjugate
optics \cite{cojugate}, measurement of atomic energy structures
and decay rates \cite{atom}. In addition, the FWM phenomenon also
has wider applications in communication networks to create new
waves and reduce loss in the signals. The easiest way to obtain
FWM in a fibre is to propagate two waves at angular frequencies
$\omega_1$ and $\omega_2$ that will create new frequencies
$\omega_3$ and $\omega_4$ such as $\omega_1 + \omega_2$ =
$\omega_3 + \omega_4$. On the other hand, the coupled version of
the above nonlocal equation, Eq.~(\ref{ablowitz model}), has also
received considerable attention and its dynamics has also  been
studied \cite{avinash}. Unlike the classical Manakov model, this
coupled nonlocal equation admits all classes of solitonic
solutions for the same nonlinearity and it does so only in the
presence of nonlocal interactions.

Motivated by the above results involving nonlocality and its unique behaviors
with wide range of applications in diverse fields, we consider the model of
coupled NLSE with nonlocal SPM and XPM along with FWM parameter. The latter is
particularly important since FWM is an essential feature of optical solitons
while carrying signals through the birefringent fibre under suitable
conditions. The details of the model are given in detail in the next section.

We employ the powerful method of Darboux transformation to find
the new solutions of the present model. We found that i) the
inclusion of FWM in the nonlocal NLSE leads to conversion of a
soliton-soliton pair , namely Bright(B)-Dark(D), B-B, or D-D, into
an Akhmediev breather, and ii) the manipulation of FWM parameters
gives a chance to observe a new solution of a {\it
breather-soliton} pair. In addition, we also find some interesting
exact solutions including the Peregrine, Kuznetsov-Ma breather,
Akhmediev breather, and a {\it breathing travelling solitonic
wave}, we called it as Soliton (Breather) lattice.

The plan of the paper is as follows. In section II, we present the
mathematical (integrable) model governing the dynamics of nonlocal
$\cal{PT}$ symmetric coupled NLSE with nonlocal FWM. In section
III, we present the corresponding Lax pair and then derive the
explicit soliton solution for zero and non-zero seed. In section
IV, we investigate the impact of FWM in nonlocal CNLSE. In section
V, we derive the special cases of the Peregrine, Ma, Akhmediev
breathers. The results are then summarized in section VI.

\section{Model Equations}
We consider the generalized $\cal{PT}$ symmetric nonlocal NLSE
with nonlocal SPM, XPM and FWM of the following form,

\begin{subequations}
\begin{align}
\textit{i}\,\frac{\partial q_{1}(x,t)}{\partial
t}+\frac{\partial^2 q_{1}(x,t)}{\partial x^2}\,+ \Big[&a\,
q_{1}(x,t)\,q_{1}^{\ast}(-x,t)\,+\, b\, q_{1}(x,t)\,
q_{2}^{\ast}(-x,t)\,\nonumber\\&+\, c\, q_{2}(x,t)
q_{2}^{\ast}(-x,t)\,+\, d\,
q_{1}^{\ast}(-x,t)\,q_{2}(x,t)\Big] q_{1}(x,t)=0,\\\notag\\
\textit{i}\,\frac{\partial q_{2}(x,t)}{\partial
t}+\frac{\partial^2 q_{2}(x,t)}{\partial x^2}\,+ \Big[&a\,
q_{1}(x,t)\,q_{1}^{\ast}(-x,t)\,+\, b\, q_{1}(x,t)\,
q_{2}^{\ast}(-x,t)\,\nonumber\\&+\, c\, q_{2}(x,t)
q_{2}^{\ast}(-x,t)\,+\, d\, q_{1}^{\ast}(-x,t)\,q_{2}(x,t)\Big]
q_{2}(x,t)=0,
\end{align}\label{fwmmodel}
\end{subequations}
where, $q_{1,2}(x,t)$ are two complex field variables and the
coefficients $a,\,c$ correspond to the nonlocal SPM and XPM while
$b,\;d$ represent the nonlocal FWM terms. The subscripts $x,t$
denote the derivatives with respect to the spatial and temporal
variables. For the sack of simplicity, the local interactions in
the above coupled NLSE $[|q_1|^2 +|q_2|^2] q_1$ and $[|q_1|^2
+|q_2|^2] q_2$ is replaced by their $\cal{PT}$ symmetric
counterparts, namely, $[q_{1} q_{1}^{\ast}(-x,t) +q_{2}
q_{2}^{\ast}(-x,t)] q_1$ and $[q_{1} q_{1}^{\ast}(-x,t) +q_{2}
q_{2}^{\ast}(-x,t)] q_2$,  along with nonlocal FWM interactions to
obtain generalized coupled nonlocal NLSE given by
Eqs.~\eqref{fwmmodel} by using the transformation given by
Eqs.~\eqref{inducenonlocal}.

\section{Lax-Pair and Darboux transformation}
\subsection{Lax-Pair}

Applying the Darboux transformation (DT) \cite{DT} method on
nonlocal generalized CNLS equation requires finding a linear
system of equations for an auxiliary fields ${\bf \Phi}(x,t)$. The
linear system is usually written in compact form in terms of the
pair of matrices as follows
\begin{subequations}
	\begin{eqnarray}
	{\bf \Phi}_{x}&=&{\bf U  \Phi},  \label{Phix} \\
	{\bf \Phi}_{t}&=&{\bf V  \Phi},  \label{Phit}
	\end{eqnarray}\label{laxcondition1}
\end{subequations}
where, ${\bf U}$ and ${\bf V}$, known as the Lax pair, are
functionals of the solutions of the model equations. The
consistency condition of the linear system ${\bf \Phi}_{xt}={\bf
	\Phi}_{tx}$ must be equivalent to the model equation under
consideration.

We find the following linear system which corresponds to the class
of generalized nonlocal coupled NLS with cross-phase and
self-phase modulation,
\begin{subequations}\label{laxcondition}
	\begin{align}
		{\bf \Phi}_{x}& = {\bf U}_{0}\,{\bf \Phi} + {\bf U}_{1}\,{\bf
			\Phi}\,{\bf \Lambda}  \\
		{\bf \Phi}_{t}& = {\bf V}_{0}\,{\bf \Phi} + {\bf V}_{1}\,{\bf
			\Phi}\,{\bf \Lambda} + {\bf V}_{2}\,{\bf \Phi}\,{\bf \Lambda}^{2}
	\end{align}
\end{subequations}
where,
%\begin{widetext}

\begin{align}
	{\bf \Phi}=
	\begin{pmatrix}
		\psi_1(x,t) & \psi_2(x,t) & \psi_3(x,t) \\\\
		\phi_1(x,t) & \phi_2(x,t) & \phi_3(x,t) \\\\
		\chi_1(x,t) & \chi_2(x,t) & \chi_3(x,t)
	\end{pmatrix},
	\;\;\; {\bf U}_{0}=\begin{pmatrix}
0 & -1 & 0 \\\\
0 & 0 & -1%
\end{pmatrix},\notag
%\label{U}
\end{align}
%\end{widetext}
%\begin{widetext}
\begin{align}
{\bf \Lambda}& =
\begin{pmatrix}
\lambda_1 & 0 & 0 \\\\
0 & \lambda_2 & 0 \\\\
0 & 0 & \lambda_3
\end{pmatrix},
\;\;\; {\bf V}_{0}=\frac{\mathrm{i}}{2}
\begin{pmatrix}
q_{1}(x,t) r_{1}(x,t)+ q_{2}(x,t) r_{2}(x,t)  & q_{1x}(x,t) & q_{2x}(x,t)
\\\\
r_{1x}(x,t) & -q_{1}(x,t) r_{1}(x,t) & -q_{2}(x,t) r_{1}(x,t) \\\\
r_{2x}(x,t) & -q_{1}(x,t) r_{2}(x,t) & -q_{2}(x,t) r_{2}(x,t)%
\end{pmatrix}, \notag
\end{align}
%\end{widetext}
%\begin{widetext}
\begin{align}
{\bf V}_{1} &=
\begin{pmatrix}
0 & -q_{1}(x,t) & -q_{2}(x,t) \\\\
r_{1}(x,t) & 0 & 0 \\\\
r_{2}(x,t) & 0 & 0%
\end{pmatrix},
\;\;\; {\bf V}_{2}=\mathrm{i}
\begin{pmatrix}
1 & 0 & 0 \\\\
0 & -1 & 0 \\\\
0 & 0 & -1%
\end{pmatrix}, \notag
%\label{V}
\end{align}\notag\\
%\end{widetext}
along with the transformation on the complex conjugates
\begin{subequations}
\begin{eqnarray}
r_1(x,t) &= a q_{1}^{\ast}(-x,t) + b q_{2}^{\ast}(-x,t)\\
r_2(x,t) &= d q_{1}^{\ast}(-x,t) + c q_{2}^{\ast}(-x,t)
\end{eqnarray} \label{inducenonlocal}
\end{subequations}
where $\lambda_{1,2,3}$ is the spectral parameter. The consistency
condition ${\bf \Phi}_{xt}={\bf \Phi}_{tx}$ leads to $ {\bf
U}_{t}-{\bf V}_{x}+[{\bf U},{\bf V}]={\bf 0}$ which should
generate the model equation \eqref{fwmmodel}.

Using the DT we have solved the above model equations using
trivial (zero) seed to obtain a single soliton solution and
non-zero seed to obtain the higher order soliton solutions.

\subsection{Darboux Transformation with zero seed: Single soliton solution}
Considering the following version of DT \cite{DT}
\begin{align}
{\bf \Phi[{\rm1}]} = {\bf \Phi  \Lambda - \sigma \Phi},
\end{align}
where, ${\bf \Phi}[1]$ is the transformed field and $\sigma =
\Phi_0\,\Lambda\,\Phi_0^{-1}$ and  ${\bf \Phi_0}$ is a known
solution of the linear system \eqref{laxcondition}, we apply the
DT on the linear system given by Eqs.~(\ref{laxcondition1}), and
the stipulation that the transformed linear system  be covariant
with the original one requires
\begin{equation}
{\bf U_0}[1]={\bf U_0}+[{\bf U_1},{\bf \sigma}] \label{comp100}.
\end{equation}
The new solution to the nonlinear equations given by,
Eqs.~(\ref{fwmmodel}), in terms of the seed solution is obtained
from the last equation.

Following this procedure, we derive the simple first order soliton
solution,
\begin{subequations}
\begin{eqnarray}
q_1(x,t) &=& \frac{A\, c\, C_1\,  e^{-(2\,\mathrm{i}\, t + 2\,
x})}{ 2\, c\, C_2\, e^{2(2\,\mathrm{i}\, t + 2\, x)} +\, C_3
(\delta_1 +
\sqrt{\delta_2})\,e^{2(2\,\mathrm{i}\, t + 2\, x)}} \\\nonumber\\
q_2(x,t) &=& \frac{A\, c\, C_1\, (\delta_1\, + \sqrt{\delta_2})
\,e^{-(2\mathrm{i}\, t + 2\, x)}}{ 2\, c\, C_2\,
e^{2(2\mathrm{i}\, t \,+\, 2\, x)} + C_3 \,(\delta_1\, +
\sqrt{\delta_2})\, e^{2(2\,\mathrm{i}\, t\, + 2\, x)}}
\end{eqnarray}\label{onesolitonsolution}
\end{subequations}
with $A$ a complex constant and $C_{1,2,3}$ are arbitrary real
constants with, $\delta_1=-b-d $,\,$\delta_2= b^2\,- 4\, a\, c\, +
2\, b\, d\, +\, d^2$. It should be noted that in order to obtain
such a localized solution we have set the following values to the
spectral parameters: $\lambda_1=\lambda_3= -\textit{i}$ and
$\lambda_2=\textit{i}\lambda_3$.
%C-3=C1, C-6=C2,C-9=C3

We observe from Eqs.~(\ref{onesolitonsolution}) that the nonlocal
FWM alongwith nonlocal SPM and XPM merely varies the amplitudes of
the solutions of the nonlocal CNLSE. Hence, we focus on the impact
of FWM alongwith SPM and XPM with nontrivial seed in the next
section.
%By inspection, we found the impact of the nonlocal FWM along with nonlocal SPM
%and XPM to be a mere change in amplitudes of the solutions of the nonlocal
%CNLSE which is a trivial effect. Therefore, we decided to focus instead on the
%FWM impact on higher order solutions, as detailed in the next section.

\subsection{Darboux Transformation with Constant-Wave seed:
Breathers and soliton pairs} Here we apply the DT using a nonzero
seed, namely the so-called Constant-Wave (CW) solution
\begin{subequations}
\begin{eqnarray}
q_1(x,t) &=& \alpha_1 \textit{e}^{\textit{i}\, \phi\, t}, \\
q_2(x,t) &=& \alpha_2 \textit{e}^{\textit{i}\, \phi\,  t},
\end{eqnarray}
\end{subequations}
where $\phi=a \alpha_1^2 + d \alpha_2^2 + 2 b \alpha_1 \alpha_2$.

The derivation of the new solution is lengthy but straightforward. We show here
only the final results. The general soliton solutions, in this case, have the
following form
\begin{eqnarray}\label{soliton-formula}
    q_k[1]=\alpha_k\left[1+\frac{{\displaystyle \sum_{l=1}^{2}
    \sum_{m=1}^{2}\frac{c_{1,l}^*c_{1,m}{\rm e}^{\theta_{1,l}^*(-x,t)
    +\theta_{1,m}(x,t)}}{\xi_{1,m}}+c_{1,3}\alpha_k\sum_{l=1}^{2}c_{1,l}^*
    {\rm e}^{\theta_{1,l}^*(-x,t)}}}{{\displaystyle -\sum_{l=1}^{2}\sum_{m=1}^{2}
    \frac{c_{1,l}^*c_{1,m}{\rm e}^{\theta_{1,l}^*(-x,t)+\theta_{1,m}(x,t)}}
    {\xi_{1,l}^*+\xi_{1,m}}+\frac{|c_{1,3}|^2B}{2(\lambda_1^*+\lambda_1)}}}
    \right]{\rm e}^{{\rm i}\phi t},
\end{eqnarray}
where
\begin{subequations}
\begin{eqnarray}
\theta_{i,1}(x,t)&=&{\rm i}\xi_{i,1}\left(x+\frac{1}{2}\xi_{i,1}t\right),\,\,\theta_{i,2}(x,t)={\rm i}\xi_{i,2}\left(x+\frac{1}{2}\xi_{i,2}t\right),\\
\beta_1&=&-(b\alpha_{{1}}+d\alpha_{{2}}-{\rm i}c\alpha_{{1}}),\,\, \beta_2=a\alpha_{{1}}+b\alpha_{{2}}+{\rm i}c\alpha_{{2}},\\
\xi_{i,1}&=&\lambda_i+\sqrt{\phi+\lambda_i^2},\,\,\xi_{i,2}=\lambda_i-\sqrt{\phi+\lambda_i^2}.
\end{eqnarray}
\end{subequations}
with $B=a|\beta_1|^2+d|\beta_2|^2+(b+{\rm i}c)\beta_1\beta_2^*+(b-{\rm
i}c)\beta_1^*\beta_2$.
\\
This solution corresponds to a family of solitonic solutions
including breathers. Each member of the family of solutions is
obtained for specific values of the parameters. In the following,
we present a detailed analysis of the nature and dynamics of these
solutions.

\section{ Impact of FWM on solutions of the nonlocal coupled NLSE}
In this section, we show the effect of FWM on the solutions of the
nonlocal CNLSE. In Sec.~\ref{sub1}, we show how FWM converts a
two-soliton pair into an Akhmediev breather. In Sec.~\ref{sub2},
we show that FWM supports new family of solutions such as a pair
of breather and a soliton and a breathing solitary wave, in
addition to the previously all known breathers including Akmediev
and Kuznetsov-Ma breathers and the Peregrine soliton.

\subsection{Two soliton solution}
\label{sub1} With no FWM, the higher order solitonic solutions are
shown to be a pair of B-B, D-D, or B-D solitons  \cite{cnsns}. For
completeness, and to show the effect of FWM, we present these
solutions here, which may be obtained from
Eq.~\eqref{soliton-formula} by choosing $c_{1,3}=0$ and
$c_{1,1}c_{1,2}\neq 0$, rendering Eq.\eqref{soliton-formula} into
the following compact form:
\begin{eqnarray}\label{soliton1}
    q_k[1]=\alpha_k\left[\frac{\cosh(\varphi_{1r})\cosh(A-2{\rm i}\varphi_{1i})+\cos(\varphi_{1i})\cosh(B+2\varphi_{1r})}{\cosh(\varphi_{1r})\cosh(A)+\cos(\varphi_{1i})\cosh(B)}\right]{\rm e}^{{\rm i}\phi t}
\end{eqnarray}
where
\begin{subequations}
\begin{eqnarray}
    %\begin{split}
       A&=&2{\rm i}\sqrt{-\phi}\sinh(\varphi_{1r})\cos(\varphi_{1i})x+\phi\cosh(2\varphi_{1r})\sin(2\varphi_{1i})t+a_1,  \\
       B&=&2\sqrt{-\phi}\cosh(\varphi_{1r})\sin(\varphi_{1i})x+{\rm i}\phi\sinh(2\varphi_{1r})\cos(2\varphi_{1i})t-{\rm i}b_1,
    %\end{split}
\end{eqnarray}
\end{subequations}
and $\exp{[(a_1+\varphi_{1r})+{\rm
i}(b_1+\varphi_{1i})]}=c_{1,1}/c_{1,2}$.  When
$\sqrt{-\phi}<\lambda_1<\sqrt{\phi}$, or $\varphi_{1r}=0$,
$0<\varphi_{1i}<\pi$, $b_1\neq (2k+1)\pi$, the solutions
\eqref{soliton1} correspond to the two-soliton solution. Along the
line
\begin{equation*}
       2\sqrt{-\phi}\sin(\varphi_{1i})x=\phi\sin(2\varphi_{1i})t+a_1\pm \ln\left(\frac{1}{|\cos(\varphi_{1i})|}\right),\,\,
       t\rightarrow \mp \infty
\end{equation*}
the height of soliton $|q_k[1]|^2$ is
\begin{equation*}
    H_1=|\alpha_k|^2\left[1+\frac{2\sin \left(\varphi_{1i}\right)
    \sin \left( b_{1}-\varphi_{1i}\right) }{1+\cos \left( b_{{1}} \right)
    }\right],
\end{equation*}
while along the trajectory
\begin{equation*}
       2\sqrt{-\phi}\sin(\varphi_{1i})x=-\left[\phi\sin(2\varphi_{1i})t+a_1\right]\pm \ln\left(\frac{1}{|\cos(\varphi_{1i})|}\right),\,\,
       t\rightarrow \pm \infty
\end{equation*}
the height of soliton $|q_k[1]|^2$ is
\begin{equation*}
    H_2=|\alpha_k|^2\left[1-\frac{2\sin \left(\varphi_{1i}\right)\sin \left( b_{1}+\varphi_{1i}\right) }{1+\cos\left( b_{1}\right)}\right].
\end{equation*}
There are three different kinds of solutions: If $H_1,H_2>|\alpha_k|^2$, it is
two-bright (B-B) soliton. If $H_1,H_2<|\alpha_k|^2$, it is two-dark (D-D)
soliton. If $H_1>|\alpha_k|^2$, $H_2<|\alpha_k|^2$ ($H_2>|\alpha_k|^2$,
$H_1<|\alpha_k|^2$), it is bright-dark (B-D) soliton, where
$\alpha_1,\beta_1,a_1,b_1$ are real arbitrary parameters and $H_{1,2}$
represents the heights of components $q_{1,2}$ respectively.

\begin{figure}[!ht]
\centering
\includegraphics[width=0.5\linewidth]{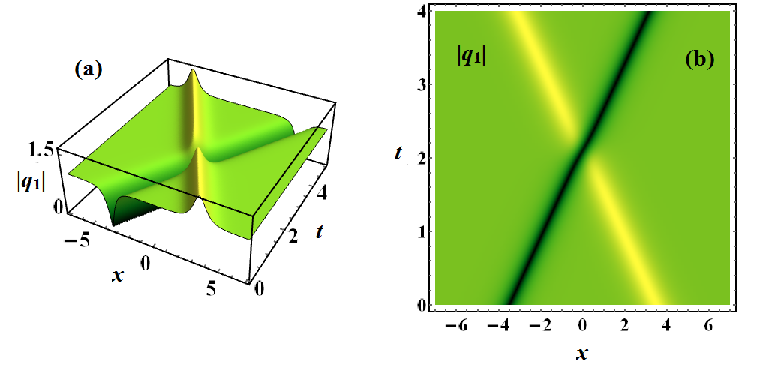}
\caption{Bright-Dark soliton given by Eq.~(\ref{soliton1}) without
FWM for the parameters $\alpha_1=1$,
$\alpha_2=\frac{1}{2}$,$a_1=8$, $b_1=2.5$, $a=c=-4,$ $b=d=0$,
$\varphi_{1r}=0$ and $\varphi_{1i}=0.2 \pi$. Similar profile
occurs also for $q_2$ component (not shown here).}\label{fwmimp1}
\end{figure}
\begin{figure}[!ht]
\centering
\includegraphics[width=0.5\linewidth]{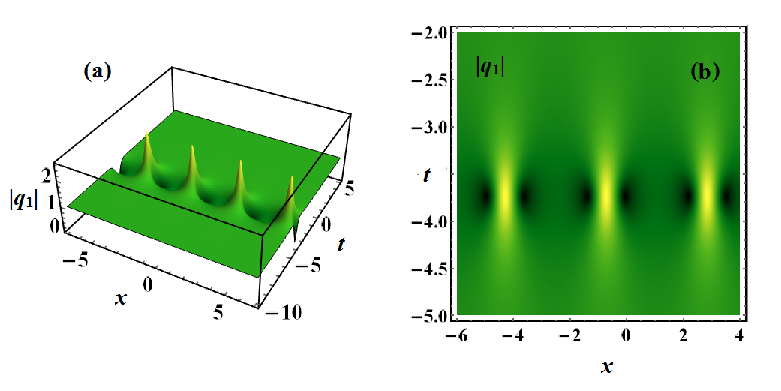}
\caption{Impact of FWM with $b$=$d$ =5 in  Eq.~(\ref{soliton1})
leading to the conversion of B-D soliton of Fig.~\ref{fwmimp1}
into breathers. Similar Profile occurs also for $q_2$ component
(not shown here).}\label{fwmimp2}
\end{figure}
\begin{figure}[!ht]
\centering
\includegraphics[width=0.5\linewidth]{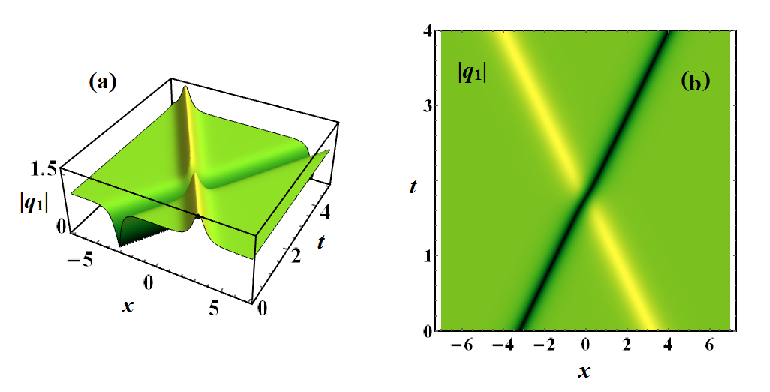}
\caption{Impact of FWM with $b\,\neq\,d$, namely $b=-5$, $d=5$ in
Eq.~(\ref{soliton1}) converting the breathers in
fig.~\ref{fwmimp2} into a B-D Soliton. Similar Profile occurs also
for $q_2$ component (not shown here).}\label{fwmimp4}
\end{figure}
\begin{figure}[!ht]
\centering
\includegraphics[width=0.5\linewidth]{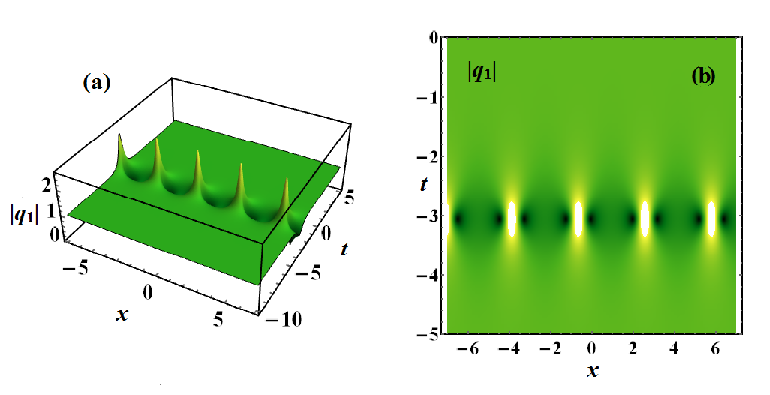}
\caption{Retrieval of breathers from B-D solitons of
Fig.~\ref{fwmimp4} by changing the signs of the FWM parameters as
$b=5$ and $d=-5$  given by Eq.~(\ref{soliton1}). Similar Profile
occurs also for $q_2$ component (not shown here).}\label{fwmimp5}
\end{figure}
\subsection{Two-Soliton-Breather Conversion}
The effect of the FWM on the group of B-B, D-D, and B-D solitons
is predominant on the B-D soliton. Introducing FWM renders the B-D
soliton into an Akhmediev breather. To understand the impact of
FWM parameters $b$ and $d$ on the nonlocal coupled NLSE along with
$a,\, c$, we, initially we take the B-D soliton without FWM
($b\,\&\,d=0$) alongwith $a=c$, as shown in Fig.~\ref{fwmimp1}.
Introducing the FWM through $b$ and $d$ namely, $b=d=5$, we obtain
the remarkable conversion of B-D solitons into Akhmediev breather,
as shown in Fig.~\ref{fwmimp2}. The manipulation of $a$ and $c$ in
Fig.~\ref{fwmimp2} with for instance, $a$ $\textless$ $c$, $a$
$\textgreater$ $c$, or $a$ $\neq$ $c$), results only in the
compression of breathers along with marginal shift towards
positive or negative time scale depending upon the choice $a$ and
$c$. Choosing FWM parameters $b$ and $d$ unequally such as $b$=-5,
$d$=5, the breather in Fig.~\ref{fwmimp2} will reverse to B-D
soliton, as shown in Fig.~\ref{fwmimp4}. One can also retrieve the
breathers, shown in Fig.~\ref{fwmimp5}, by interchanging the signs
between $b$ and $d$ like 5 and -5, respectively, as shown in
Fig.~\ref{fwmimp4}. The right panels are the corresponding density
plots for the left panels with reduced time and space axes for
better view. Another interesting impact of FWM on the nonlocal
coupled NLSE is that it destroys the sensitivity to the parameter
$b_1$. In other words, we found earlier  \cite{cnsns} that the
manipulation of the different classes of soliton solutions such as
B-B, B-D, D-B, D-D solitons can be performed  sensitively by the
parameter $b_1$. The inclusion of FWM parameters in the model
makes the free parameter $b_1$ a passive member in the group of
parameters, which means the change of $b_1$ arbitrarily will not
affect the density of the profiles shown in
Fig.~\ref{fwmimp1}-\ref{fwmimp5} in any sense, as long as FWM
parameters $\neq 0$. Instead of $b_1$, the FWM parameters $b\,\&
d$ play an active role in the conversion of any type of solitons
like B-B,B-D,D-B,D-D to breathers and viceversa. We would like to
add that wherever the two components exhibit similar behavior with
a mere change in amplitude, we have plotted only one component
($q_1$) while we plot two components ($q_1$ and $q_2$) when they
exhibit different behavior.

%{\bf It is must to mention here that, we have shown only one
%component $(q_1)$ when the two components exhibit similar profiles
%with a mere change in amplitude whereas we showed both two
%components $(q_1$ and $q2)$ where they are completely different
%from each other.}

\section{Family of higher order soltions with FWM}
\label{sub2} Here, we present a family of unique solitonic
solutions and breathers. Some of these solutions are obtained only
with FWM such as the breather-soliton solution in,
Fig.~\ref{fig2}, solitonic waves in, Fig.~\ref{fig3}, and a new
breather that is periodic in both time and space axes, as shown in
Fig.~\ref{fwmspl}d, which we denote here as ``Soliton (Breather)
lattice".
\subsection{Breather-Soliton solution}
\begin{figure}[!ht]
\centering
\includegraphics[width=0.5\linewidth]{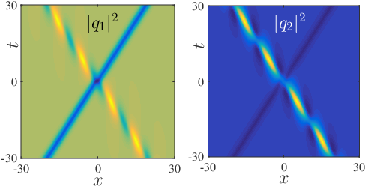}
\caption{(color online): Breather-Soliton solution given by
Eq.~(\ref{soliton1}): bright soliton-breather (left) and dark
soliton-breather (right) for the choice of parameters:
$\alpha_1=1$, $\alpha_2=\frac{1}{4}$, $a=c=-1,$ $b=1$, $d=1$,
$c_{1,1}=c_{1,2}^{-1}={\rm e}^{\frac{{\rm i}\pi}{4}}$,
$c_{1,3}=1$, $\varphi_{1r}=0$ and $\varphi_{1i}=\frac{\pi}{4}$
}\label{fig2}
\end{figure}

The breather-soliton solution is obtained for
$c_{1,1}c_{1,2}c_{1,3}\neq 0$,
$\exp{\frac{1}{2}[(a_1+\varphi_{1r})+{\rm
i}(b_1+\varphi_{1i})]}=c_{1,1}$ and\\
$\exp{-\frac{1}{2}[(a_1+\varphi_{1r})+{\rm
i}(b_1+\varphi_{1i})]}=c_{1,2}$. Since the non-singularity
condition for this type of solution is very complex, we merely
give a sufficient condition
\begin{equation*}
  \begin{split}
     \cos(\varphi_{1i})\sin(b_1)>0,&  \\
     1+\frac{|c_{1,3}|2}{4}|B|+\cos(\varphi_{1i})\cos(b_1)>0.&
  \end{split}
\end{equation*}
In this breather-soliton solution, one dark soliton or bright
soliton is replaced by the breather, as shown in Fig.~\ref{fig2}.
%\cite{recent}

\subsection{Solitonic waves}
In this section, we present a unique solitonic wave which is a
travelling wave with time-dependent amplitude, as shown in
Fig.~\ref{fig3}, and is obtained by a careful choice of
parameters, as given below. When we choose
$|\lambda_1|>\sqrt{-\phi}$, $\lambda_1\in \mathbb{R}$, or
$\varphi_{1r}\in \mathbb{R}$, $\varphi_{1i}=0$: There are three
kinds of periodic solutions.

If $c_{1,3}=0$, the solution \eqref{soliton-formula} can be
rewritten as
\begin{eqnarray}
\label{periodic}
    q_k[1]=\alpha_k\left[\frac{\sinh(\varphi_{1r})\sinh(A-2{\rm i}\varphi_{1i})-\cos(\varphi_{1i})\cosh(B+2\varphi_{1r})}{\sinh(\varphi_{1r})\sinh(A)+\cos(\varphi_{1i})\cosh(B)}\right]{\rm e}^{{\rm i}\phi t}
\end{eqnarray}
where
\begin{eqnarray}
       A&=&2{\rm i}\sqrt{\phi}\cosh(\varphi_{1r})\cos(\varphi_{1i})x-\phi\cosh(2\varphi_{1r})\sin(2\varphi_{1i})t+a_1,  \\
       B&=&2\sqrt{\phi}\sinh(\varphi_{1r})\sin(\varphi_{1i})x-{\rm i}\phi\sinh(2\varphi_{1r})\cos(2\varphi_{1i})t-{\rm i}b_1,
\end{eqnarray}
and $\exp{[(a_1+\varphi_{1r})+{\rm
i}(b_1+\varphi_{1i})]}=c_{1,1}/c_{1,2}$. Here, we consider a special
case $\varphi_{1r}\in \mathbb{R}/\{0\}$, $\varphi_{1i}=0$. When
$|\sinh(\varphi_{1r})\sinh(a_1)|>1$, then one can obtain periodic
solution. Similar to the defocusing case, we can obtain another
periodic solution when $c_{1,3}\neq 0.$ We ignore here the analysis
of the non-singularity condition since it is very complex.

If $c_{1,2}=0$, $c_{1,1}=1$, $c_{1,3}\neq 0$ and
$2\cosh(\varphi_{1r})\neq {\rm e}^{\varphi_{1r}}|c_{1,3}|^2|B|$,
or $c_{1,1}=0$ and $c_{1,2}c_{1,3}\neq 0$, then solution
\eqref{soliton-formula} is also a periodic solution.

We have shown periodic solution here in Fig.~\ref{fig3} for the
choice $c_{1,1}=0$, $c_{1,2}c_{1,3}\neq 0$ given by
Eq.~(\ref{periodic}).
\begin{figure}[!ht]
\centering
\includegraphics[width=0.5\linewidth]{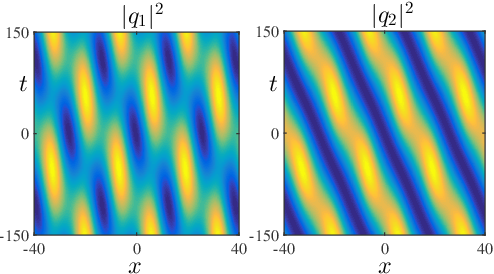}
\caption{(color online): Periodic solutions given by
Eq.~(\ref{periodic}) for the choice of parameters $\alpha_1=1$,
$\alpha_2=\frac{1}{4}$, $a=c=-1,$ $b=\frac{1}{2}$,
$d=\frac{1}{2}$, $c_{1,1}=0$, $c_{1,2}={\rm e}^{\frac{-{\rm
i}\pi}{4}}$, $c_{1,3}=1$, $\varphi_{1r}=\mathrm{arccosh}(2)$ and
$\varphi_{1i}=0$.}\label{fig3}
\end{figure}
\subsection{Breathers: Kuznetsov-Ma breather, Akhmediev
breather, and Peregrine soliton} 
\begin{figure}[!ht]
\centering
\includegraphics[width=0.4\linewidth]{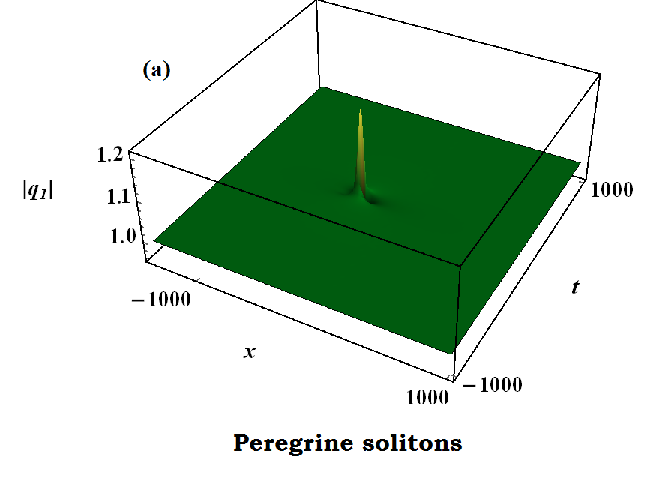}
\includegraphics[width=0.4\linewidth]{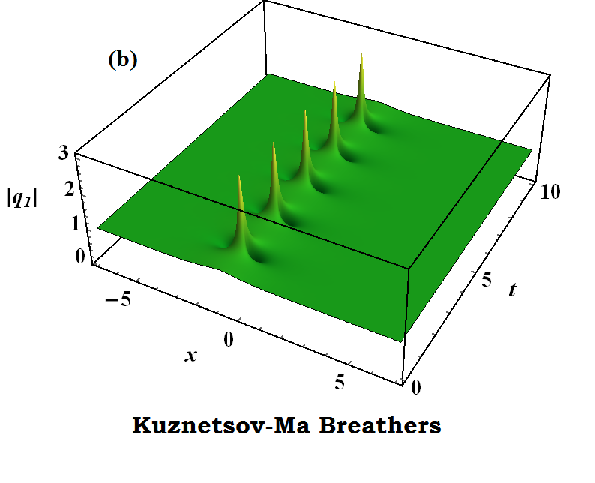}\\
\includegraphics[width=0.4\linewidth]{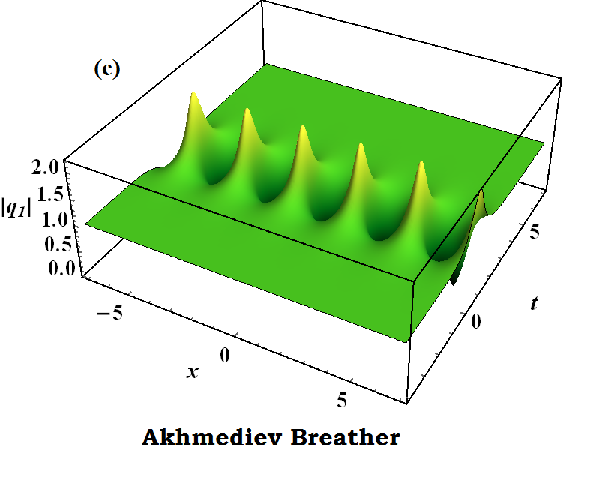}
\includegraphics[width=0.4\linewidth]{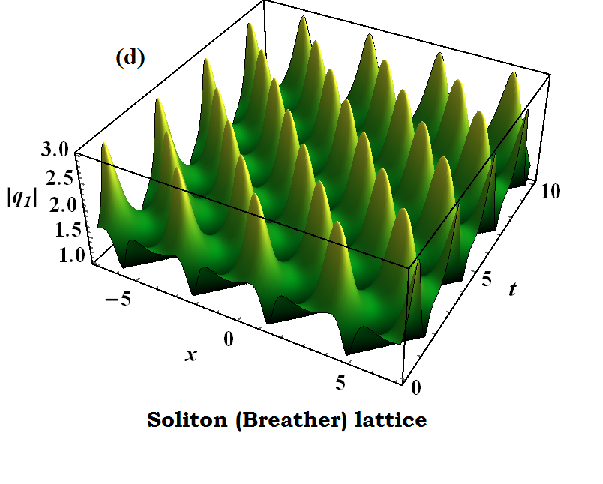}
\caption{Different types of breathers given by
Eq.~(\ref{periodic}): (a) Peregrine solitons for the choice of
parameters $\alpha_1=1$, $\alpha_2=\frac{1}{2}$, $a=c=-1,$ $b=5$,
$d=5$, $a_1=0$, $b_1=2.5$ $\varphi_{1r}=0$ and
$\varphi_{1i}=0.00013 \pi$, (b) Kuznetsov-Ma breathers for the
choice of parameters $\varphi_{1r}=0.3$ and $\varphi_{1i}=0$, with
$a_1=-1$, $b_1=1$ other parameters are same as in (a).~(c)
Akhmediev breather for the choice of parameters $a=c=-4,$ $b=5$,
$d=5$, $a_1=-2$, $b_1=2.5$, $\varphi_{1r}=0$ and $\varphi_{1i}=0.3
\pi$, (d) Soliton (breather) lattice for the choice of parameters
$a=c=-1,$ $b=0.1$, $d=0.1$, $a_1=0$, $b_1=1$ $\varphi_{1r}=0.8$
and $\varphi_{1i}=0$.}\label{fwmspl}
\end{figure}

In addition to the above effect
of FWM, we found that the nonlocal coupled NLSE given by
Eq.~(\ref{fwmmodel}) supports some interesting profiles such as
Peregrine solitons, Kuznetsov-Ma breather, Akhmediev breathers,
Space-Time breather \cite{kuznetsov}. Peregrine solitons are
obtained for the choice of parameters such that the oscillatory
and exponential terms in the solution given by,
Eq.~(\ref{periodic}) vanish each other. This is obtained by the
choice $\varphi_{1r}=0$ and $\varphi_{1i} \neq 0$, as shown in
Fig.~\ref{fwmspl}a. For the choice of parameter
$\varphi_{1r}\neq0$ and $\varphi_{1i}=0$, one obtains spatially
localized temporally periodic Kuznetsov-Ma breather as shown in
Fig.~\ref{fwmspl}b. One can also obtain temporally localized and
spatially periodical Akhmediev breathers as shown in
Fig.~\ref{fwmspl}c for the choice of parameter $\varphi_{1r}=0$
and $\varphi_{1i}=0.3 \pi$. We also notice an interesting profile
which is both spatially and temporally periodic which can be also
viewed as a travelling breathing wave along the time axis with a
phase shift. We called it as Soliton (breather) lattice, as shown
in Fig.~\ref{fwmspl}d. We believe that these breathers are found
for the first time in the literature for the model involving
nonlocal interactions with FWM.

\section{Conclusion}
In this paper, we investigated $\cal{PT}$ symmetric coupled
nonlinear schr\"odinger equation with nonlocal four-wave mixing
employing Darboux transformation to generate a family of exact
solitonic solutions including the Peregrine soliton, Kuznetsov
Ma-breather, Akhmediev breather, and a breather which is periodic
in both space and time (we named it them Soliton (breather)
lattice). In addition, two soliton solutions including
bright-bright, dark-dark, and bright-dark turn out to be solutions
of the model. We have shown that the inclusion of four-wave mixing
in the nonlocal nonlinear schr\"odinger equation leads to
conversion of bright-dark solitons into breathers. We also observe
that by manipulating the four-wave mixing parameters associated
with other arbitrary parameters of the system, one can observe all
possible conversions between soliton-soliton interaction into
breathers and vice versa. We believe that the above phenomenon
occurs due to the nonlocal nature of the dynamical system with
nonlocal four-wave mixing.

\section*{Acknowledgements}
UAK and PSV acknowledge the support of UAE University through the
grant UAEU-UPAR(7) and UAEU-UPAR(4). RR wishes to acknowledge the
financial assistance received from Department of Atomic
Energy-National Board for Higher Mathematics (DAE-NBHM) (No.
NBHM/R.P.16/2014) and Council of Scientific and Industrial
Research (CSIR) (No. 03(1323)/14/EMR-II) for the financial support
in the form Major Research Projects. LL acknowledge the financial
support received from National Natural Science Foundation of China
(Contact No. 11401221).


\begin{thebibliography}{99}
\bibitem{hasegawa} Hasegawa A, Tappert F. Transmission of stationary nonlinear optical pulses in dispersive dielectric fibers. I. Anomalous
dispersion. Appl Phys Lett 1973; {\bf 23}: 142.


\bibitem{manakov1}
Manakov S V. On the theory of two dimensional statioanary
self-focusing of electro magnetic waves. Zh Eksp Teor Fiz 1973;
{\bf 65}: 505.

\bibitem{agrawal} Agrawal G P. Nonlinear Fibre optics. New york: Academic
Press; 2006.


\bibitem{pre-psv} Radha R, Vinayagam P S, Porsezian K.
Rotation of the trajectories of bright solitons and realignment of
intensity distribution in the coupled nonlinear Schr\"odinger
equation. Phys Rev E 2013; {\bf 88}: 032903.

\bibitem{manakov2}
Kaup D J,  Malomed B A. Soliton trapping and daughter waves in the
Manakov model. Phys Rev A 1993; {\bf 48}: 599.

\bibitem{modifiedmanakov}
Makhankov V G, Makhaldiani N V, Pashaev O K. On the integrability
and iso-otpic structure of the one dimensional Hubbard model in
the long wave approximation. Phys Lett A 1981; {\bf 81}: 161.


\bibitem{nlo}
Park Q H, Shin H J. Systematic construction of vector soltions.
IEEE J Quantum Electron 2002; {\bf 8}: 432.

\bibitem{bio}
Scott A C. Launching a Davydov Soliton: I. Soliton Analysis. Phys
Scr 1984; {\bf 29}: No 3: 279.

\bibitem{finance}
Yan Z. Vector financial rogue waves. Phys Lett A 2011; {\bf 375}:
4274.

\bibitem{oceano}
Dhar A K, Dhas K P. Fourth-order nonlinear evolution equation for
two Stokes wave trains in deep water. Phys Fluids A 1991; {\bf 3}:
3021.

\bibitem{ablow} Ablowitz M J, Musslimani Z H. Integrable Nonlocal Nonlinear Schrödinger Equation. Phys Rev Lett 2013; {\bf 110}: 064105.


\bibitem{aksarma} Sarma A K, Miri M A, Musslimani Z H, Christodoulides D N. Continuous and discrete Schrödinger systems with parity-time-symmetric
nonlinearities. Phys Rev E 2014; {\bf 89}: 052918.


\bibitem{exp1} Musslimani Z H, Makris K G, El-Ganainy R,
Christodoulides C N. Optical Solitons in PT PT Periodic
Potentials. Phys Rev Lett 2008; {\bf 100}: 030402.

\bibitem{exp2} Bludov Yu V, Konotop V V, Malomed B A. Stable dark solitons in PT-symmetric dual-core waveguides. Phys Rev A
 2013; {\bf 87}: 013816.

\bibitem{exp3} Driben R, Malomed B A. Stability of solitons in parity-time-symmetric couplers. Opt Lett 2011; {\bf 36}: 4323.

\bibitem{exp4} Miri M-A, Aceves A B, Kottos T, Kovanis V, Christodoulides D N. Bragg solitons in nonlinear PT-symmetric periodic
potentials. Phys Rev A 2012; {\bf 86}: 033801.


\bibitem{plasmonics} Benisty H, Degiron A, Lupu A, De Lustrac A, Chenais S, Forget S, Besbes M, Barbillon G, Bruyant A, Blaize S, Lerondel G. Implementation of PT symmetric devices using plasmonics: principle
and applications. Opt Express 2011; {\bf 19}: 18004.

\bibitem{opt-meta-1} Kulishov M, Laniel J, Belanger N, Azana J, Plant D. Nonreciprocal waveguide Bragg gratings. Opt Express 2005;
{\bf 13}: 3068.

\bibitem{opt-meta-2} Lin Z, Ramezani H, Eichelkraut T, Kottos T, Cao H, Christodoulides D N. Unidirectional Invisibility Induced by PT-Symmetric Periodic Structures. Phys Rev Lett 2011; {\bf
106}: 213901.

\bibitem{cohe-atom} Sheng J, Miri M-A, Christodoulides D N, Xiao M. PT-symmetric optical potentials in a coherent atomic medium. Phys
Rev A 2013; {\bf 88}: 041803(R).


\bibitem{yang} Yang J. Nonlinear Waves in Integrable and Nonintegrable
Systems. SIAM: 2010.
\newline
Wang D S, Zhang D J, Yang J. Integrable properties of the general
coupled nonlinear Schrödinger equations. 2010; {\bf 51}: 023510.

\bibitem{opt-proc}
Pepper D M, AuYeung J, Fekete D and  Yariv A. Spatial convolution
and correlation of optical fields via degenerate four-wave mixing.
Opt. Lett. 1978; {\bf 3}: 7.

\bibitem{holo}
Gerritsen H J. Nonlinear effects in image formation. Appl. Phys.
Lett 1967; {\bf10}: 239.

\bibitem{cojugate}
Yariv A. Quantum Electronics.New York: John Wiley \& Sons;1989.

\bibitem{atom}
Yajima T and Souma H. Study of ultra-fast relaxation processes by
resonant Rayleigh-type optical mixing. I. Theory. Phys. Rev. A
1978; {\bf17}: 309.

\bibitem{avinash} Khare A, Saxena A. Periodic and Hyperbolic soliton solutions of a number of nonlocal PT-symmetric nonlinear equation.
arXiv;1405.5267.

\bibitem{DT} Matveev V B, Salle M A. Darboux Transformations and Solitons. Berlin:  Springer-Verlag;1991.

\bibitem{cnsns} Vinayagam P S, Radha R, Khawaja U Al, Liming L. Collisional dynamics of solitons in the coupled PT symmetric nonlocal NLS
equation. Commun Nonlinear Sci Numer Simulat. 2017; {\bf 52}: 1.

\bibitem{kuznetsov} Kibler B, Fatome J, Finot C, Millot G, Genty G, Wetzel B, Akhmediev N, Dias F, Dudley J M. Observation of Kuznetsov-Ma soliton dynamics in optical fibres. Sci Rep. 2012;
{\bf 463}: 2.

%\bibitem{recent} Song C Q, Xiao D-M, Zhu Z-N. Solitons and
%dynamics for a general integrable nonlocal coupled nonlinear
%schrodinger equation. Com Non Sci Num Sim 2017; {\bf 45}: 13.

\end{thebibliography}
\end{document}